# Signal and Noise Analysis in TRION -Time-Resolved Integrative Optical Fast Neutron Detector

D. Vartsky[a]*, G. Feldman[a], I. Mor[a], M. B. Goldberg[a], D. Bar[a], V. Dangendorf[b]

[a] *Soreq NRC,*
*Yavne 81800, Israel*
[b] *Physikalisch-Technische Bundesanstalt (PTB),*
*38116 Braunschweig, Germany*
*E-mail*: *david@soreq.gov.il*

ABSTRACT: TRION is a sub-mm spatial resolution fast neutron imaging detector, which employs an integrative optical time-of-flight technique. The detector was developed for fast neutron resonance radiography, a method capable of detecting a broad range of conventional and improvised explosives. In this study we have analyzed in detail, using Monte-Carlo calculations and experimentally determined parameters, all the processes that influence the signal and noise in the TRION detector. In contrast to event-counting detectors where the signal-to-noise ratio is dependent only on the number of detected events (quantum noise), in an energy-integrating detector additional factors, such as the fluctuations in imparted energy, number of photoelectrons, system gain and other factors will contribute to the noise. The excess noise factor (over the quantum noise) due to these processes was 4.3, 2.7, 2.1, 1.9 and 1.9 for incident neutron energies of 2, 4, 7.5, 10 and 14 MeV, respectively. It is shown that, even under ideal light collection conditions, a fast neutron detection system operating in an integrative mode cannot be quantum-noise-limited due to the relatively large variance in the imparted proton energy and the resulting scintillation light distributions.

KEYWORDS: Fast neutron detectors, Scintillating fiber screen, Neutron radiography, Time-of-flight spectroscopy, Detection of explosives, Detective quantum efficiency, Excess noise factor.

# Contents

# 1. Introduction

The TRION detector was developed for fast neutron resonance radiography (FNRR) [1], a fast-neutron transmission imaging method that exploits characteristic energy-variations of the total scattering cross-section in the $E_n$ = 1-10 MeV range to detect specific elements within a radiographed object. FNRR holds promise for detecting a broad range of conventional and improvised explosives, due to its ability to determine simultaneously the identity and density distribution of the principal elements present in explosives, such as C, O and N.

The variant of FNRR with a pulsed neutron beam, known as Pulsed Fast Neutron Transmission Spectroscopy (**PFNTS**), was proposed and first studied by the Oregon University group [2,3,4] for detection of explosives. The method was subsequently refined and taken through several blind tests for the FAA by Tensor-Technology, Inc. [5,6,7], In the PFNTS method, a ns-pulsed, broad-energy (1-10 MeV) neutron beam is incident on the inspected object and the transmitted neutron spectrum measured by the Time-of-Flight (TOF) technique. Both the Oregon University and Tensor groups employed the conventional event counting TOF (**ECTOF**) spectroscopy mode, in which the elapsed time for an individual neutron to arrive at the detector following its creation in the target during the beam burst is recorded.

Both groups employed large-area detector arrays consisting of individual plastic scintillators (dimensions: several cm), each coupled to a photomultiplier tube via a light guide. The pixel size determined by these detectors posed an intrinsic limitation on the position resolution, which did not permit reliable detection of small and thin objects. Reduction of pixel size while using the above approach would entail an increase in the quantity of electronics at a prohibitive cost.

TRION is a sub-mm spatial resolution fast neutron detector, which employs integrative optical TOF (**ITOF**) technique. As opposed to **ECTOF** it integrates the detector signal during a well-defined gate time at a pre-selected $t_{TOF}$ corresponding to a pre-selected energy bin, e.g., the energy-interval spanning a cross-section resonance. The TOF spectrum is obtained by varying the delay of the integrating gate relative to the time of the burst [8,9,10]. In an integrative detector the quantum information such as the energy deposited by each detected event and its exact arrival time is lost and only the integral information is recorded. The advantages of this approach are: 1) it can provide excellent spatial resolution at an affordable cost and 2) it permits operation at unlimited neutron fluxes. The disadvantages are: 1) TOF resolution is dictated by gate width and detector response; 2) Sequential scanning is required for accumulation of entire time spectrum and 3) a reduced signal-to-noise ratio.

In comparison to event-counting detectors where the signal-to-noise ratio is dependent only on the number of detected events, in an energy-integrating detector additional factors such as the fluctuations in imparted energy, number of photoelectrons, system gain and other factors will contribute to the noise. Analyses of the signal-to-noise ratio in integrative imaging radiation detectors have been performed by several investigators. Swank [11] analyzed the effect in X-ray phosphors and introduced a factor, which arises from fluctuations in the number of light photons emitted from the screen per absorbed X-ray. Other investigators [12-17] studied the signal-to-noise-ratio in various X-ray screens coupled to optical detectors and in storage phosphors [18].

Lanza et al [19] calculated the variance in signal of a cooled, CCD-based thermal neutron imaging system. Barmakov et al [20] investigated the detection quantum efficiency and spatial resolution of thermal neutron detectors based on $^6$LiF+ZnS:Ag and $Gd_2O_2S$:Tb screens together with a CCD camera. Mikerov et al [21] described a physical model of fast neutron interaction in thin (2 mm) disperse screens consisting of a hydrogenous polymer matrix with suspended powder luminophore, such as ZnS:Ag and $Gd_2O_2S$:Tb and also in 2-10 mm thick transparent screens made of polymethylmethacrylate. Using Monte-Carlo calculations they calculated the detection quantum efficiency, luminosity and inherent spatial resolution of the screens.

In this work we analyze the propagation of the signal produced by the various components of the TRION fast neutron detector and study the physical processes within them that influence the signal-to-noise ratio.

## 2. Description of TRION detector

### 2.1 Principle of operation of TRION

A detailed description of TRION can be found in [22]. Here we shall present only a schematic description of the system with emphasis on system components that have a major influence on the signal-to-noise ratio. Fig. 1 illustrates the detector principle.

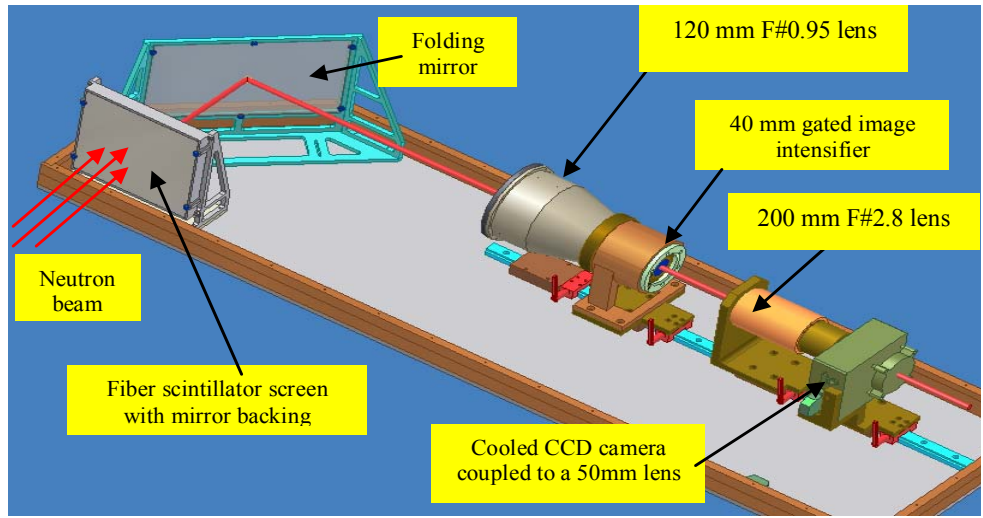

Fig. 1 Description of TRION detector

A broad energy spectrum of fast neutrons is produced by an accelerator in 1-2 ns bursts with a frequency of 2 MHz. After passing through an inspected object, these neutrons arrive at the detector after their time-of-flight, interact in a scintillating plastic fiber screen and produce a light image. The scintillation light transport optics consists of a front-surface, good-quality mirror positioned at 45° to the neutron beam axis and a large-aperture 120 mm F#0.95 lens that transfers the image to a cooled, gated image-intensifier (I-I), which acts as an electronic shutter that is opened for a gate period $\Delta t$ at a selected $t_{TOF}$ (or neutron energy).

A cooled CCD camera views the image created at the phosphor of the gated intensifier via a lens relay. Within the accelerator pulsing period of 500 ns, depending on the neutron source-detector distance and the width of the relevant energy bin, the detector integrates neutrons into an image in a well-defined TOF-bin relative to the beam pulse. By varying the gate delay time $t_{TOF}$, transmission images at any selected neutron energy can be taken.

#### 2.1.1 Generation of signal in TRION

Fig. 2 shows the sequence of signal generation in TRION.

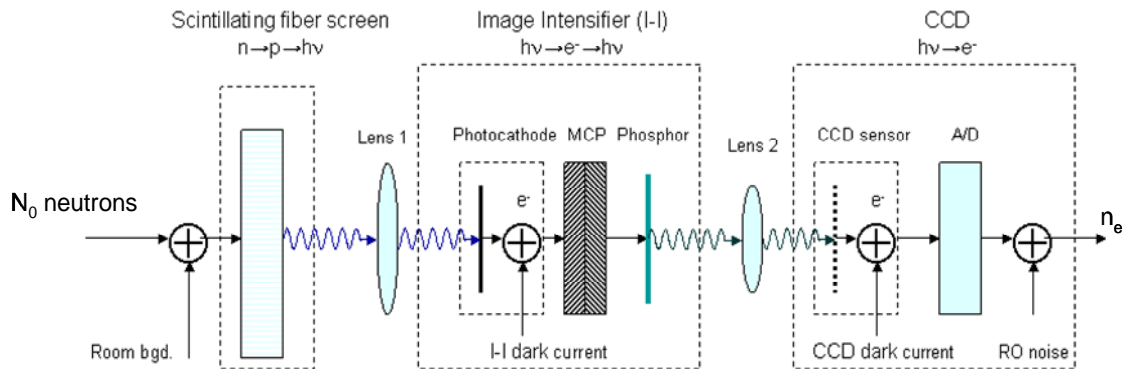

**Fig.2: Schematic description of the signal sequence developed in TRION**

The physical processes that govern the signal generated on the CCD sensor per incident neutron in TRION are:
1. The interaction of the incident neutron in the scintillating screen, resulting in transfer of part of its energy to a proton
2. The conversion of energy dissipated by the proton in the scintillator to light
3. The transport of light photons to the photocathode of the image intensifier through lens-1
4. The absorption of light photons at the photocathode and the emission of photoelectrons
5. The electron multiplication process within the image intensifier and its re-conversion to light at its exit phosphor.
6. The transport of intensified light from the image intensifier to the faceplate of the CCD through lens 2.
7. The absorption of light photons by the CCD sensor and its conversion to electrons.

The average number of electrons generated on the CCD sensor by $N_0$ <u>*detected*</u> neutrons of energy $E_n$, can be expressed (see A4.1 in the Appendix) as:

$$\mathbf{n_e = N_0 \cdot Ph(E_n) \cdot T_1 \cdot p_1 \cdot m_G \cdot T_2 \cdot p_2} \qquad Eq.1$$

where,

$N_0$ - number of neutrons detected over the exposure time

$Ph(E_n)$ - average number of light photons emitted from the scintillating fiber per detected neutron of energy $E_n$

$T_1$ - transmission of lens 1

$p_1$ - quantum efficiency of I-I

$m_G$ - average gain of I-I (photons/electron)

$T_2$ - transmission of lens 2

$p_2$ - quantum efficiency of CCD

In addition to the signal generated by neutrons, additional signals enter the chain at various stages. These are the dark currents of I-I and CCD due to single electrons generated by thermal energy in the I-I photocathode and in the CCD Si matrix and the CCD readout noise (**RN**). If during the measurement time we detect $N_0$ neutrons and generate $N_1$ and $N_2$ dark current electrons in the I-I and CCD respectively, the average overall measured signal will be:

$$\mathbf{n_{eT} = N_0 \cdot Ph(E_n) \cdot T_1 \cdot p_1 \cdot m_G \cdot T_2 \cdot p_2 + N_1 \cdot m_G \cdot T_2 \cdot p_2 + N_2 + RN} \qquad Eq.2$$

In order to obtain the net neutron signal the dark signals and RN must be determined separately and subtracted from the overall signal. In TRION we reduce the dark current signals to insignificant levels by cooling the I-I and the CCD.

### 2.1.2 The noise in TRION

In the Appendix we show (Eq. A8.3) that the relative noise (reciprocal of signal to noise ratio) in an integrated signal (due to $N_0$ detected neutrons), resulting from a cascade of **m** processes can be expressed as:

$$\frac{\sigma_{n_{eT}}}{n_{eT}} = \frac{1}{\sqrt{N_0}} \left[ 1 + \sum_m \left\{ \left(\frac{\sigma_m}{m_m}\right)^2 \left(\frac{M_m}{n_{eT}}\right) \right\} \right]^{1/2} \qquad Eq.3$$

where $m_m$ and $\sigma_m$ are the mean and standard deviation of a probability distribution function (**PDF**) of the process "**m**" and $M_m$ is the product of the means $N_0 \cdot m_1 \cdot m_2 \cdot m_3 \cdot \ldots m_m$.

Thus in an integrating system such as TRION, the relative noise will always be larger than the relative quantum noise $1/\sqrt{N_0}$ of the detected incident signal by an excess noise factor (**ENF**) that depends on **weighted** relative errors of all physical processes that contribute to formation of the signal. The inverse square of this factor divided by the detection efficiency is also referred to as **Detective Quantum Efficiency (DQE)**, it signifies how well the detector reproduces the signal to noise ratio (SNR) of the incident radiation [23].

In TRION the main contributors to this excess noise are the following processes:
1. Generation of light photons by neutrons in the screen and transport to I-I
2. Creation of photoelectrons on the I-I photocathode (binomial process)
3. Amplification of light in I-I and transport to CCD sensor
4. Creation of photoelectrons in the CCD sensor (binomial process)

The magnitude of the mean value $m_m$ and the relative error $\sigma_m/m_m$ of each process is dependent on its PDF.

We shall now describe in detail the properties of the scintillating screen, the optical system and the image intensifier and determine the PDF of the processes contributing to the ENF.

## 3. Determination of PDF and its parameters for each process

The PDF of the above processes were determined either experimentally or by Monte Carlo simulation.

### 3.1 Generation of light photons by neutrons in the screen and transport to I-I

#### 3.1.1 Description of the fiber screen

The scintillating fiber screen is manufactured by Saint Gobain (formerly Bicron), USA [24]. Its surface area is 200×200 mm². The fiber screen is made of 30 mm long scintillating fibers, each consisting of a 0.5×0.5 mm² polystyrene core (refractive index-$n_1$=1.60), a 20 μm thick Poly-Methyl-MethAcrylate (PMMA) cladding (refractive index-$n_2$=1.49), and a 16 μm thick, $TiO_2$ doped, white polyurethane paint coating, which acts as Extra Mural Absorber-(EMA) to prevent light cross-talk. The fibers were first assembled in 10×10 mm² square bundles, which were then glued together to form the entire screen. The front face of the screen is covered with a high quality mirror that reflects the component of light travelling backward along the fiber towards the light collecting system. Fig. 3 shows the scintillating fiber screen and a magnified view of a section of the scintillating fiber screen.

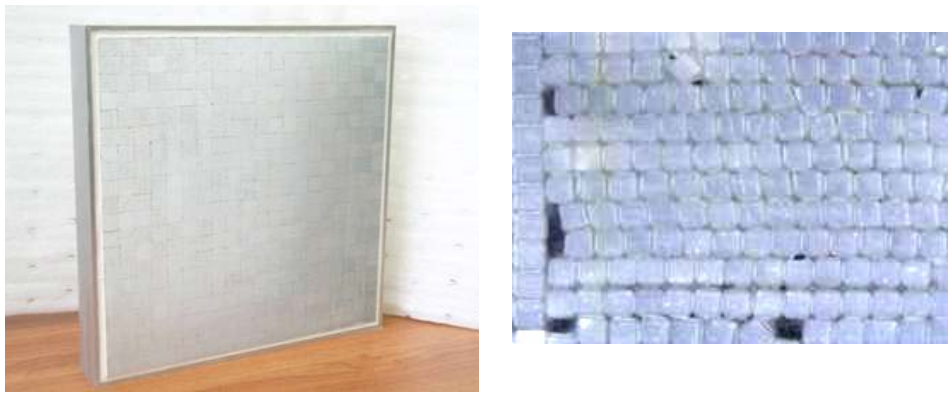

**Fig. 3. Entire scintillating fiber screen (left) and magnified view of one region (right)**

As can be observed, the screen is by no means uniform and has many imperfections. The local net sensitive area was evaluated for different regions of the screen and ranged between 59% and 72%. The overall net sensitive area is 63%.

### 3.1.2 Detailed simulation of the response of the scintillating fiber screen

#### 3.1.2.1 Proton energy spectra

In order to understand the contribution of screen response to the signal-to-noise ratio, it is important to determine the distribution of the light output for mono-energetic neutrons.

A neutron entering the screen can transfer its energy to a knock-on proton according to the angle between the scattered proton and the incident neutron:

$$E_p = E_n \cos^2(\theta) \qquad Eq.\ 4$$

where $E_p$ is proton energy, $E_n$ is the energy of the incident neutron and $\theta$ is the proton scattering angle in the lab coordinate system. Thus the knock-on proton can acquire energies ranging from zero to the full neutron energy. As the neutron-proton reaction is isotropic throughout the energy range of interest here (1-14 MeV), the energy distribution of the protons is flat. Thus, even for mono-energetic neutrons, the distribution of proton energies is very broad. The situation is further modified by the presence of carbon in the screen and by the fact that the screen is composed of fibers of which only the core is active in creating light. The cladding and EMA paint will absorb part of the proton energy, but will not generate a scintillation signal.

Some of the knock-on protons created in the core may escape it, leaving only part of their energy inside the core. On the other hand, protons created outside the core may enter it and deposit their energy there. Thus the energy distribution of protons capable of creating light in the fiber core is expected to deviate from a pure flat distribution.

A detailed calculation of the distribution of energy deposited in a scintillating fiber core has been performed using the GEANT 3.21 code. The simulated setup consisted of a 200×200×30 mm$^3$ fiber screen, uniformly irradiated at 5 different neutron energies (2, 4, 7.5, 10 and 14 MeV) by a 200×200 mm$^2$ mono-energetic parallel-beam of neutrons impinging on the screen face. Fig. 4 shows a schematic configuration of 9 fibers located in the center of the screen and a magnified view of the central fiber (0.5×0.5 mm$^2$ polystyrene core, 20 μm thick PMMA cladding and 16 μm thick EMA paint).

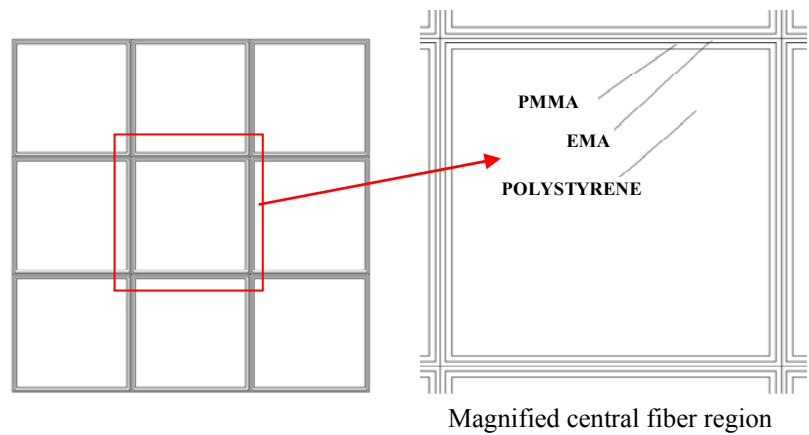

Magnified central fiber region

**Fig. 4. Schematic of 9-fiber array (left) and magnified view of central fiber (right)**

The simulation calculates the energy deposited by protons in the core of the central fiber (test fiber). For tracking and tallying purposes, protons created by incident neutrons entering the central fiber are termed "*primary protons*". Protons created outside the central fiber that reach its core are referred to as "*secondary protons*" and protons created by neutrons entering the central fiber after being scattered into it from any of the outer regions are defined as ***"tertiary protons"***.

For a given neutron energy, the number of secondary protons depends on the proton range and that of the tertiary protons depends on neutron beam dimensions and screen geometry. Since the tertiary protons are created by neutrons scattered within the screen, their contribution will vary with position across the screen.

In the simulation, the same number of neutrons ($6\times10^9$) uniformly incident on the $200\times200$ mm$^2$ area was employed at each of the 5 neutron energies.

Fig. 5 shows the distribution of energy deposited by the primary, secondary and tertiary protons in the central fiber at 3 incident neutron energies (2, 7.5 and 14 MeV).

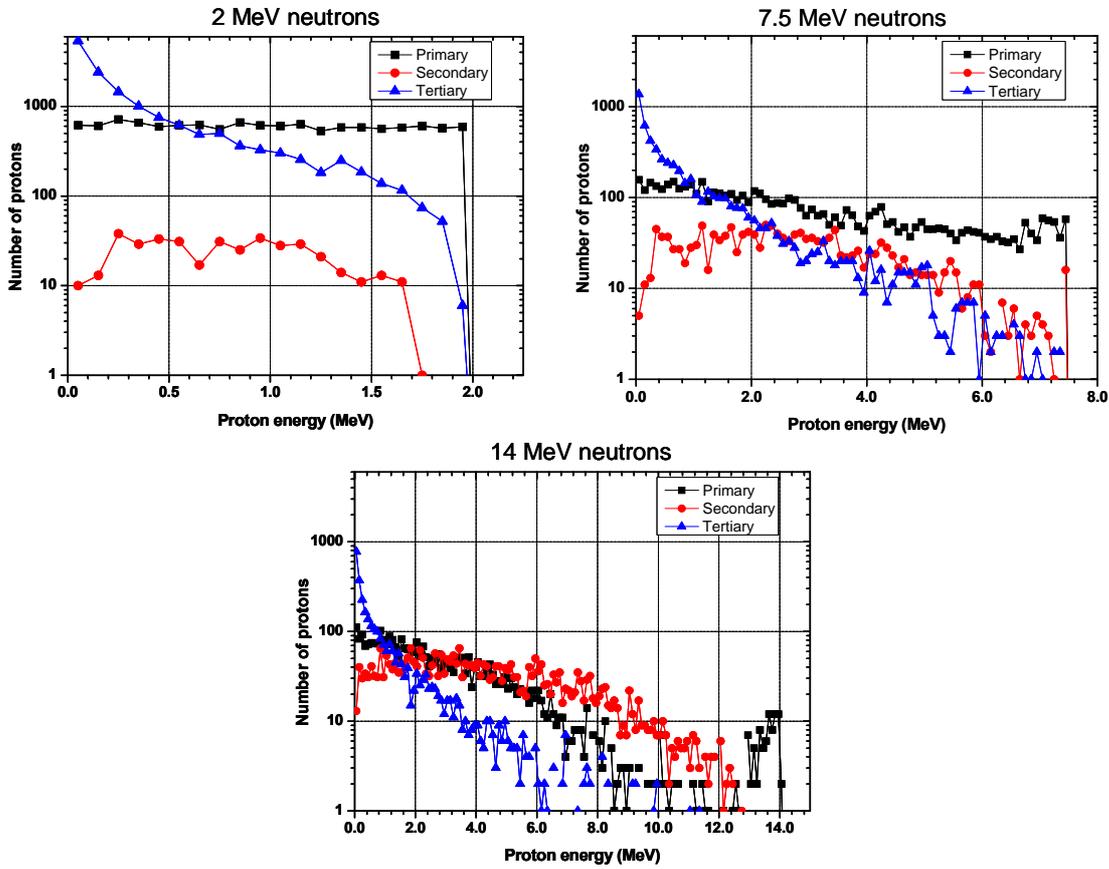

**Fig. 5 Energy deposited in the central fiber by primary (black), secondary (red) and tertiary (blue) protons for 2 MeV (top left), 7.5 MeV (top right) and 14 MeV (bottom) neutrons**

As can be observed, with increasing neutron energy, the distribution of the primary protons deviates from the flat distribution. The reason for this is that, as the proton energy increases, the proton range increases accordingly and if the track crosses a pixel boundary it will deposit only part of its energy in the fiber core in question. The secondary proton energy distribution extends over a wide energy range and that of the tertiary protons is skewed toward lower energies.

Table 1 shows the contributions of primary, secondary and tertiary protons to the total number of protons created in the fiber core.

**Table 1 Contribution and mean energy of primary, secondary and tertiary protons**

| Neutron energy (MeV) | Primary protons (%) | | Secondary protons (%) | | Tertiary protons (%) | |
|---|---|---|---|---|---|---|
| | Contribution (%) | Mean energy | Contribution (%) | Mean energy | Contribution (%) | Mean energy |
| 2 | 44 | 0.98 | 1.5 | 0.78 | 54.5 | 0.36 |
| 4 | 45 | 1.82 | 4 | 1.38 | 51 | 0.67 |
| 7.5 | 43 | 2.80 | 13 | 2.73 | 44 | 0.92 |
| 10 | 39 | 3.2 | 25 | 3.6 | 36 | 1.0 |
| 14 | 35 | 2.95 | 34 | 4.3 | 31 | 1.0 |

As can be observed a large fraction of protons reaching the fiber core originate from neutrons that interacted outside the test fiber. The proportion of secondary protons increases with neutron energy due to the increase in proton range with its energy. Thus there is a larger chance for protons created in the neighbourhood of the test pixel to reach it and deposit part of their energy in its core. The proportion of the tertiary protons is relatively large, however the average energy that they deposit is low.

The total mean proton energy deposited in the fiber is 0.64 MeV, 1.21 MeV, 1.96 MeV, 2.5 MeV and 2.8 MeV for the 2 MeV, 4 MeV, 7.5 MeV 10 MeV and 14 MeV neutrons respectively. This is significantly lower than half the neutron energy, the value usually taken to represent the average proton energy.

### 3.1.2.2 Light intensity distribution

The scintillation light intensity distribution does not follow exactly the proton energy distribution because of the non-linear behaviour of scintillation light generation with proton energy. The response of plastic scintillator to protons has been studied by several investigators [25]. The response is usually expressed in electron equivalent energy ($MeV_{ee}$), i.e the electron energy which would produce the same amount of light output as that produced by a proton of energy $E_p$. Fig. 6 shows the amount of light $L(E_p)$ vs proton energy $E_p$ in plastic scintillators obtained using experimental $MeV_{ee}$ data of O'Reilly et al [25] together with the specific light yield of 8000 photons/$MeV_{ee}$ provided by Saint Gobain [24]. A polynomial function fit to this data is also shown on the figure.

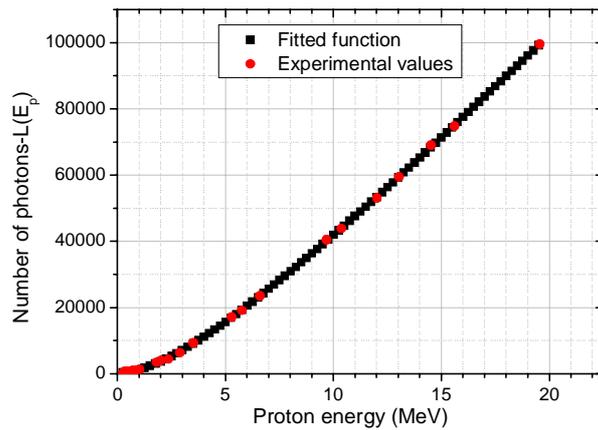

**Fig. 6 Light output $L(E_p)$ of scintillating plastic fiber vs proton energy $E_p$**

As can be observed L(E$_p$) exhibits non-linear behaviour with E$_p$ especially below 5 MeV. The amount of light **L** created in the core of a fiber by a proton can be calculated by:

$$L = \int_{E_f}^{E_i} \left(\frac{dL}{dE}\right) dE = L(E_i) - L(E_f) \qquad Eq.\ 5$$

where E$_i$ and E$_f$ are the initial and final energy respectively of a proton traveling within the fiber core. The values of E$_i$ and E$_f$ are obtained by the Monte-Carlo calculation for each proton. The above non-linear behaviour will cause further skewness in the light distribution toward low light emission.

Only a fraction of the amount of light **L** is transmitted to the end of the fiber and is emitted toward the collecting lens. This fraction (trapping efficiency) is dependent on the refractive indices of core and cladding. According to the manufacturer data [24] the trapping efficiency of single-clad square fibers is **4.4%**, i.e. only 4.4% of the total light created in the scintillation travels within the fiber to each end of the fiber. The white EMA coating decreases the amount of light obtained from the fiber, because the coating can interfere with light transmission in the cladding. The light output of a fiber with white EMA was measured to be about **65%** of that of a bare fiber.

By placing a good quality front face mirror on the front (incident beam) side of the screen face one can increase the above fraction of light exiting the fiber toward the collecting lens by a factor of 1.8 [28]. Thus the amount of light emitted from the fiber is expected to be ~**5%** of the total light created in a scintillation.

Fig. 7 shows the frequency distribution of light emitted from the fiber screen per detected neutron for 3 neutron energies (2, 7.5 and 14 MeV).

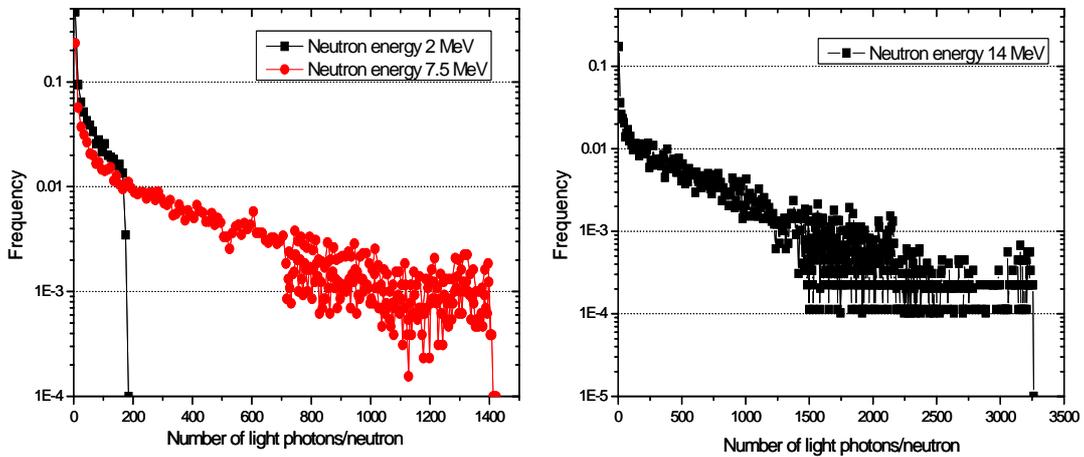

**Fig. 7 Probability distribution of light emitted from the fiber screen**

The mean number of light photons/neutron emitted from the fiber **Ph(E$_n$)** and the standard deviation of the distribution is **36.6±35.6, 113±112, 271±271 333±332 and 377±376** photons for neutron energies of 2, 4, 7.5, 10 and 14 MeV respectively.

### 3.1.2.3. Light collection efficiency

The light emission from the scintillating fiber is limited to a cone, whose apex angle $\theta$ is determined by the refractive indices $n_1$ of the core and $n_2$ of the cladding material. In our fiber screen the maximum emission angle $\theta_{fiber\,max}$ is ~35.7° [24]. The angular distribution of the light intensity within this cone is approximately constant [27].

This light is viewed by a custom made, large aperture lens (120 mm F#0.95) positioned at a distance of 750 mm from the scintillating screen. The fraction of light collected by the lens is determined by the ratio of the solid angle subtended by the lens to that defined by the fiber emission cone:

$$g = \frac{(1-\cos\theta_{lens})}{(1-\cos\theta_{max\,fiber})} \qquad Eq.\ 6$$

For a lens diameter of 126 mm and fiber-to-lens distance of 750 mm, $\theta_{lens}$= 4.8° and $\theta_{max\,fiber}$= 35.45° hence, $g$=0.019.

The transmission of our lens is about 90% so the fraction of light reaching the photocathode of the I-I is **$T_1$=0.017**.

Table 2 presents the mean number of light photons per detected neutron reaching the I-I photocathode-$m_{Ph}$=$Ph(E_n)\cdot T_1$, the standard deviation of the distribution $\sigma_{Ph}$ and the relative error of the distribution for neutron energy of 2, 4, 7.5, 10 and 14 MeV respectively.

**Table 2 Parameters of PDF for photons reaching the I-I photocathode**

| Neutron energy (MeV) | Mean number of photons/n -$m_{Ph}$ | Standard deviation $\sigma_{Ph}$ | $\sigma_{Ph}/m_{Ph}$ |
|---|---|---|---|
| 2 | 0.62 | 0.6 | 0.97 |
| 4 | 1.92 | 1.9 | 0.99 |
| 7.5 | 4.6 | 4.6 | 1.0 |
| 10 | 5.7 | 5.6 | 0.98 |
| 14 | 6.5 | 6.4 | 0.99 |

### 3.2 Generation of the signal in the image intensifier

### 3.2.1 Conversion to photoelectrons

The gated image-intensifier shown on Fig. 7 (manufactured by PHOTEK Limited, UK [28]), is a high-gain, proximity-focus device. The tube is about 25 mm in length and 40 mm in diameter with a rugged metal ceramic construction. The input window is made of fused silica. The photocathode is a low noise S20 with a conductive mesh undercoating. The manufacturer quotes the quantum efficiency at a wavelength of 420 nm to be 13%. Two micro-channel plates enable a radiant gain of $10^6$ W/W. The output window is made of fiber optic and a P43 phosphor screen.

A typical dark emission rate of a low noise S20 photocathode is about 500 e/s/cm$^2$. In order to reduce the dark noise, the I-I photocathode was cooled by blowing cold dry air on the face of the I-I fused silica window. After about 20 minutes of cooling the dark noise level is reduced by a factor of about 50. Conversion of photons into photoelectrons (ph-e) is a **binomial** process. Thus, for the mean number of photoelectrons/ photon (QE) **$p_1$**, the standard deviation of the process is $\sqrt{p_1 \cdot (1-p_1)}$. It follows that the mean number of photoelectrons/photon is **$p_1$=0.13 and $\sigma_{p_1} = 0.34$.** Thus the relative standard deviation of this distribution is $\sigma_{p_1}/p_1$=2.6.

### 3.2.2 Distribution of I-I gain and transfer of light to CCD

Following the creation of the photoelectrons in the photocathode they are transferred (by a photocathode voltage of –150 V) towards the MCP, where they are multiplied and accelerated again (under a voltage of +3900 V) toward the P43 phosphor screen.

Several factors affect the distribution of I-I gain:
- electron transfer from photocathode to MCP-in
- multiplication in MCPs
- generation of light in phosphor

The electronic I-I gain distribution has been determined by measuring the single electron pulse height distribution on the I-I phosphor. This measurement does not take into account the statistics of the light emission from the phosphor, however, since the average number of photons/electron is large we do not expect this process to affect the I-I gain distribution to any significant extent.

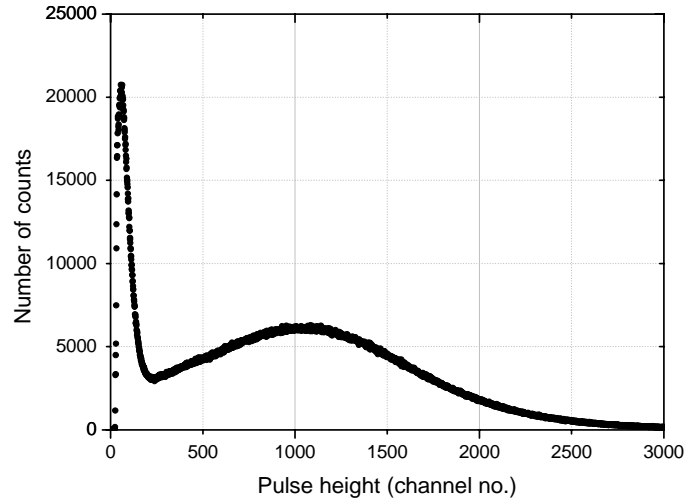

**Fig. 8 Distribution of I-I gain**

Fig. 8 shows the single electron pulse height spectrum. The average gain as quoted by the manufacturer[28] is **$m_G$=9×10$^6$** photons/electron. Only a fraction **$T_2$=0.008** of the light generated by the I-I phosphor is transmitted by the 2nd lens, (a tandem assembly consisting of Canon 200 mm F#2.8 and Nikkon 50 mm F#5.6 lenses) to the CCD sensor. Thus the average number of photons/e- at the CCD sensor is $m_G \cdot T2$=7.2×10$^4$ and the measured relative standard deviation of the above distribution is $\sigma_G/m_G$=0.68.

### 3.3 Conversion of light to electrons on CCD sensor

The final process in the cascade is the conversion of photons emitted from the I-I phosphor screen to a CCD camera electronic signal. This conversion occurs at the CCD sensor and is also a binomial process. The quantum efficiency QE of a standard CCD sensor at 550 nm is 0.**55**. It follows that the mean number of photoelectrons/photon is **$p_2$=0.55; $\sigma_{p2}$=0.49; $\sigma_{p2}/p_2$=0.89**.

## 4. Excess Noise Factor in TRION

As shown by Eq. 3, the ENF depends on **$(\sigma_m/m_m)^2 \cdot (M_m/n_{eT})$**, or the weighted squares of the relative standard deviations of all physical processes that contribute to the formation of the signal. Table 3 shows the value of the above expression for each of the 4 processes described in section 3 for 5 neutron energies

**Table 3. Value of weighted squares of the relative errors for each process and ENF**

| Neutron energy (MeV) | 1st process Generation of light in fiber screen | 2nd process Generation of ph-e on I-I photocathode | 3rd process I-I amplification | 4th process Generation of ph-e on CCD sensor | ENF |
|---|---|---|---|---|---|
| 2 | 0.97 | 10.8 | 5.74 | $1.4 \times 10^{-4}$ | 4.3 |
| 4 | 0.99 | 3.5 | 1.85 | $4.5 \times 10^{-5}$ | 2.7 |
| 7.5 | 1.0 | 1.45 | 0.8 | $1.9 \times 10^{-5}$ | 2.1 |
| 10 | 0.98 | 1.2 | 0.6 | $1.5 \times 10^{-5}$ | 1.9 |
| 14 | 0.99 | 1.1 | 0.6 | $1.4 \times 10^{-5}$ | 1.9 |

The contribution of the first process is nearly constant with energy and represents the fluctuations in imparted energy and thus in the amount of light reaching the I-I (see table 2). The contribution of the 2nd and 3rd processes depends inversely on the mean number of photoelectrons produced on the I-I photocathode per neutron and decreases with neutron energy. The contribution of the 4th process is insignificant compared to the other processes due to the large number of photoelectrons generated in the CCD sensor following the light amplification.

## 5. Discussion

The total light created in a scintillating fiber is due to energy deposited by primary, secondary and tertiary protons. Only primary protons carry spatial information of the incident beam. The contribution of the secondary protons will cause some deterioration of the spatial resolution (especially at high neutron energies) and that of tertiary protons will give rise to a diffused background that may reduce image contrast. However, as is evident from Fig. 5 and Table 1, the mean energy of tertiary protons is substantially lower than that of the other contributors, due to the fact that they originate from neutrons scattered within the fiber screen. Thus, although the number of tertiary protons is comparable to the other contributors, they create small amount of light and will add relatively little to the total number of photoelectrons created at the I-I photocathode.

TRION detector exhibits a relatively large ENF especially at low neutron energies. The main reason for this is the rather low number of photoelectrons produced at the I-I photocathode. In principle, the ENF may be improved by increasing the scintillator light yield, collection efficiency $T_1$ and quantum efficiency $p_1$ of the image intensifier. It is difficult to achieve a substantial increase in the scintillation yield of a plastic scintillator, however the two other factors can in principle be improved. Fig. 9 shows the decrease in ENF vs. increase in light collection efficiency (left) and I-I quantum efficiency (right).

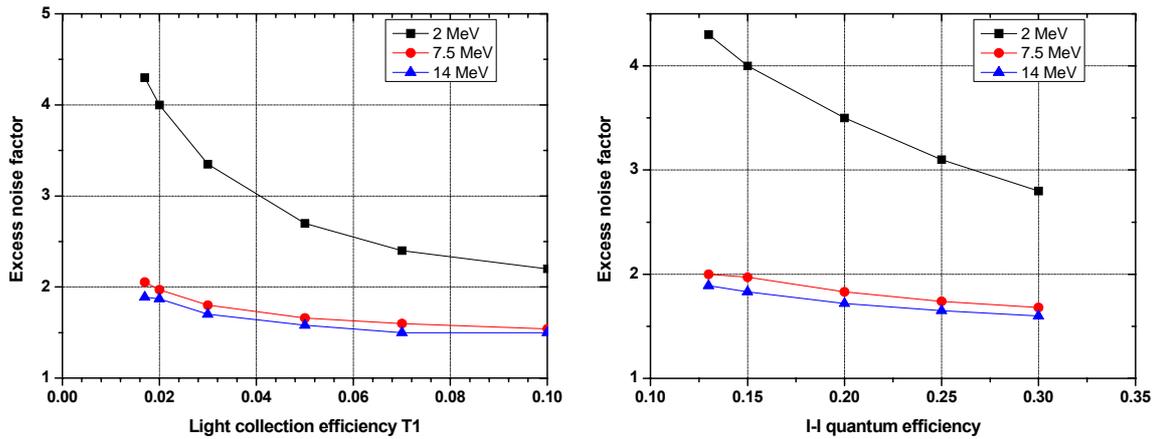

**Fig. 9 ENF vs. light collection efficiency (left) and I-I quantum efficiency**

As can be observed the ENF decreases with the increase in both factors. However, even with a very substantial increase in light magnitude at the I-I photocathode the lowest ENF achievable with TRION will be **1.41**, due to the fact that even for monoenergetic neutrons the light created by protons has a wide and skewed distribution. Thus a TRION detector cannot be quantum noise limited.

In addition to the contributors to signal variance described in previous sections there are additional sources of noise which are not dependent on the signal. These are the thermal noises of the image intensifier and the CCD and the readout noise of the CCD. In TRION we reduce the thermal noise values to insignificant levels by cooling the photocathode of the I-I and the CCD faceplate. The CCD readout noise can become significant at low neutron count rates.

As mentioned before, the TRION detection system cannot be quantum noise limited. A comparison with an event counting optical readout system is being performed at present and will be published in a separate paper.

## 6. References

[1] Vartsky D., *Prospects of Fast-Neutron Resonance Radiography and its Requirements for Instrumentation*- Review paper, in Proceedings Int. Workshop on Fast Neutron Detectors and Applications, University of Cape Town, South Africa, April 3 – 6, 2006, http://pos.sissa.it/archive/conferences/025/084/FNDA2006 084.pdf.

[2] Overley J.C., *Determination of H, C, N, O content of bulk materials from neutron-attenuation measurements*, 1985 Int. J. Appl. Rad. & Isot. **36** 185

[3] Overley J.C., *Element-Sensitive Computed Tomography with Fast Neutrons*, 1987, Nucl. Instr. and Meth. **B 24/25** 1058

[4] Overley J.C., *Explosives detection through fast-neutron time-of-flight attenuation measurements*, 1995, Nucl. Instr. and Meth. **B 99** 728

[5] Miller TG and Makky WH, *Application of Fast Neutron Spectroscopy/Radiography (FNS/R) to Airport Security*, 1992 Proceedings of the SPIE, **Vol. 1737 Neutrons, X-Rays and Gamma Rays**, pp. 184-196

[6] Miller T.G. and Krauss R.A., *Substance identification using neutron transmission*, 1995 SPIE Proceedings, **Vol. 2511** 14

[7] Miller T.G., Van Staagen P.K., Gibson B.C., Orthel J.L. and Krauss R.A., *Contraband detection using neutron transmission*, 1997 SPIE Proceedings, **Vol 2936** 102.

[8] Dangendorf V., Kersten C., Laczko G., Jaguztky O., Spillman U., *Fast Neutron Resonance Radiography in a Pulsed Neutron Beam*, in 7[th] World Conf. On Neutron Radiography, Rome, Sept. 2002, http://arxiv.org/abs/nucl-ex/0301001

[9] Dangendorf V., Kersten C., Laczko G., Vartsky D., Mor I., Goldberg M.B., Feldman G., Jaguztky O., Spillman U., Breskin A., Chechik R., *Detectors for energy resolved fast neutron imaging*, 2004 Nucl. Instr. and Meth. **A 535** 93

[10] Vartsky D., Mor I., Goldberg M.B., Mardor I., Feldman G., Bar D., Shor A., Dangendorf V., Laczko G., Breskin A. and Chechik R., *Time Resolved Fast Neutron Imaging: Simulation of Detector Performance*, 2005 Nucl. Instr. Meth. **A542** 197

[11] Swank R.K., *Absorption and noise in X-ray phosphors*, 1973 J. Appl. Phys **44** 4199

[12] Dick C.E. and Motz J.W., *Image information transfer properties of X-ray fluorescent screens*, 1981 Med. Phys. **8(3)** 337

[13] Chan. H. and Doi K., *Studies of X-ray energy absorption and quantum noise properties of X-ray screens by use of Monte Carlo simulation*, 1984 Med. Phys. **11(1)** 37

[14] Rowlands JA and Taylor KW, *Absorption and noise in cesium iodide X-ray image intensifiers*, 1983 Med. Phys. **10(6)** 786

[15] Rabbani M. and Van Metter R., *Analysis of signal and noise propagation for several imaging mechanisms*, 1989 J. Opt. Soc. Am **6(8)** 1156

[16] Sandborg M. and Carlsson G.A., *Influence of X-ray energy spectrum, contrasting detail and detector on the signal-noise ratio (SNR) and detective quantum efficiency (DQE) in projection radiography*, 1992 Phys. Med. Biol **37** 1245

[17] Ginzburg A. and Dick C.E., *Image information transfer properties of X-ray intensifying screens in the energy range from 17 to 320 keV,* 1993 Med. Phys. **20(4)** 1013

[18] Witt P., *Detective quantum efficiency of storage phosphors for soft X-rays*, 1993 Pure Appl. Opt. **2** 61

[19] Lanza R.C., Shi S., McFarland E.W., *A cooled CCD based neutron imaging system for low fluence neutron sources*, 1996 IEEE Trans. Nucl. Sci. **43** 1347

[20] Barmakov J.N., Bogolubov EP, Koshelev A.P., Mikerov V.I. and Ryzhkov V.I., 2004 Nucl. Instr. Meth. **B213** 241

[21] Mikerov V.I., Zhitnik I.A., Barmakov J.N., Bogolubov E.P., Ryzhkov VI, Koshelev A.P., Sohin N.P., Washkowski W., Lanza R.C. and Hall J.M., 2004 Appl. Rad. Isotopes **61** 529

[22] Mor I., *Energy resolved fast neutron imaging via time resolved optical readout*, M.Sc. Thesis, 2006, http://jinst.sissa.it/theses/2006_JINST_TH_002.jsp

[23] Dainty C. and Shaw R., Image Science (Academic, New York 1974)

[24] Saint-Gobain Crystals, www.detectors.saint gobain.com

[25] O'Reilly GV, Kolb NR, Pywell RE, *The response of plastic scintillator to protons and deuterons*, 1996 Nucl. Instr. Meth. **A368** 745

[26] Cohen M., 2007 Private communication

[27] Maidment ADA and Yaffe MJ, *Analysis of signal propagation in optically coupled detectors for digital mammography: II. Lens and fibre optics*, 1996 Phys. Med. Biol. **41** 475

[28] Photek Ltd., UK, www.photek.com/products/detectors_image_intensifiers

[29] http://en.wikipedia.org/wiki/Law_of_total_expectation

[30] http://en.wikipedia.org/wiki/Law_of_total_variance

[31] Thirwall J.T. *The detective quantum efficiency of medical x-ray image intensifiers*, 1998. Rev. Sci. Instrum., **69** 3953

[32] Breitenberger E. *Scintillation spectrometer statistics*, 1955 Prog. in Nucl. Phys., **4** 56

# 7. Appendix

# Noise propagation formulas for a general cascade of particle transformations

The appendix contains the self-sufficient derivation of these formulas. The principal equation (Eq. A4.2) could be found in a slightly different form in [31] and even earlier [32].

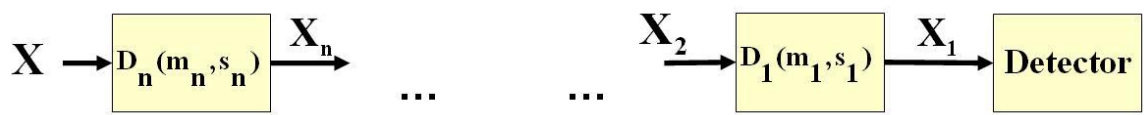

**Fig. A1 Block diagram of a cascade of particle transformations**

**1. Notations.**
Let us describe a cascade of particle transformations described by Fig. A1 by:
$$X = X_{n+1} \Rightarrow X_n \Rightarrow ... X_2 \Rightarrow X_1 = C_n(X) \qquad Eq.A1$$
where each $X_i$ is the (random!) number of particles generated at the i-th step of the cascade. We suppose that at each step
$$X_i = \sum_{j=1}^{X_{i+1}} X_{i,j} \qquad Eq.A2$$
where each $X_{i,j}$ is the random number of particles, generated by a single particle that came from the previous step. We assume that all these $X_{i,j}$ are identically distributed and are uncorrelated random values with mean $m_i$ and standard deviation $s_i$.

For each random value $Y$ we will denote as $M(Y)$ and $Var(Y)$ its mathematical expectation (mean) and its variance.

Let us denote
$$M_0 = 1, \quad M_i = m_1 \cdot m_2 \cdot ... \cdot m_i, \quad k_i = s_i / m_i, \quad i = 1,2,...,n \qquad Eq.A3$$

**2. Proposition.**
$$M(C_n(X)) = M(X) \cdot M_n \qquad Eq.A4.1$$
$$\frac{Var(C_n(X))}{M(C_n(X))^2} = k_X^2 + \frac{K_n}{M(X)}, \text{ where } k_X = \frac{Var(X)}{M(X)^2}, \quad K_n = \frac{1}{M_n} \sum_{m=1}^{n} k_m^2 \cdot M_m \qquad Eq.A4.2$$

**Proof.**
First, let's observe that, because $X_{i,j}$ are uncorrelated, conditional mean and conditional variance of the $X_n$ will equal
$$M(X_n \mid X) = \sum_{j=1}^{X} M(X_{n,j}) = m_m \cdot X \qquad Var(X_n \mid X) = \sum_{j=1}^{X_{i+1}} Var(X_{n,j}) = s_n^2 \cdot X \qquad Eq.A5$$
By application the Law of Total Expectation [29],
$$M(U) = M(M(U \mid V))$$
and the Law of Total Variance [30],
$$Var(U) = M(Var(U \mid V)) + Var(M(U \mid V))$$
to the equalities Eq. A5), we derive:
$$M(X_n) = M(m_n \cdot X) = m_n \cdot M(X) \qquad Eq.A6.1$$
$$Var(X_n) = M(s_n^2 \cdot X) + Var(m_n \cdot X) = s_n^2 \cdot M(X) + m_n^2 \cdot Var(X) \qquad Eq.A6.2$$
Now the equalities (Eq. A4.1) ,(Eq. A4.2) will be proved by induction on n. For n=0 these equalities are trivial (no multipliers in (Eq. A4.1), no sum in (Eq. A4.2). Suppose, the equalities are true for all cascades of length n-1, and for all signals X. Then it is true for the signal $X_n$, so that

$$M(C_n(X)) = M(X_n) \cdot M_{n-1} \qquad Eq.A7.1$$

$$\frac{Var(C_n(X))}{M(C_n(X))^2} = \frac{Var(X_n)}{M(X_n)^2} + \frac{1}{M(X_n)} \cdot \sum_{m=1}^{n-1} k_m^2 \cdot \frac{M_m}{M_{n-1}} \qquad Eq.A7.2$$

Substituting (Eq. A6.1) into (Eq. A7.1) we obtain (Eq. A4.1) immediately. Let's substitute (Eq. A6.1) and (Eq. A6.2) into (Eq. A7.2):

$$\frac{Var(C_n(X))}{M(C_n(X))^2} = \frac{Var(X_n)}{M(X_n)^2} + \frac{1}{M(X_n)} \cdot \sum_{m=1}^{n-1} k_m^2 \cdot \frac{M_m}{M_{n-1}} = \frac{s_n^2 \cdot M(X) + m_n^2 \cdot Var(X)}{m_n^2 \cdot M(X)^2} +$$

$$+ \frac{1}{m_n \cdot M(X)} \cdot \sum_{m=1}^{n-1} k_m^2 \cdot \frac{M_m}{M_{n-1}} = \frac{Var(X)}{M(X)^2} + \frac{1}{M(X)} \cdot k_n^2 \cdot \frac{M_n}{M_n} + \frac{1}{M(X)} \cdot \sum_{m=1}^{n-1} k_m^2 \cdot \frac{M_m}{M_n}$$

The last expression coincides with the right side of (Eq. A4.2).

**3.1. Corollary**. If the input signal $X$ is Poisson-distributed, so that $M(X) = Var(X) = N$, then

$$M(C_n(X)) = N \cdot M_n \qquad Eq.A8.1$$

$$\frac{Var(C_n(X))}{M(C_n(X))^2} = \frac{1}{N} \cdot (1 + K_n) \qquad Eq.A8.2$$

follows immediately by substitution of $M(X) = Var(X) = N$ into (Eq. A4.1), (Eq. A4.2).

Thus the relative error is $\dfrac{1}{\sqrt{N}} \cdot (1+\mathbf{K_n})^{1/2}$ or:

$$\boxed{\frac{1}{\sqrt{N}} \cdot \left(1 + \sum_{m=1}^{n} k_m^2 \cdot \frac{M_m}{M_n}\right)^{1/2}} \qquad Eq.A8.3$$

Now we could apply our formulas to the more general and more realistic scenario, including into our consideration a background input and noise output at each stage of the cascade:

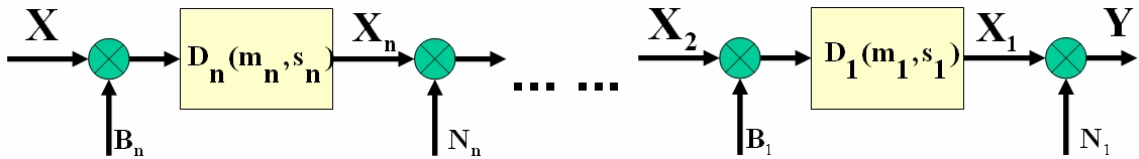

Fig. A2. General cascade

We still suppose, that all these random values are uncorrelated. For each random value Z we use notation $K_Z = Std(Z)/M(Z)$.

**3.2. Corollary**. For the cascade depicted in Fig. A2

$$M(Y) = M(X) \cdot M_n + \sum_{m=1}^{n} M_{m-1} \cdot (m_m \cdot B_m + N_m) \qquad \frac{Var(Y)}{M(C_n(X))^2} = k_X^2 + \frac{K_n}{M(X)} + \Delta,$$

$$\Delta = \sum_{m=1}^{n} \left[ \left(\frac{M(B_m)}{M(X)} \cdot \frac{M_m}{M_n}\right)^2 \cdot \left(k_{B_m}^2 + \frac{K_m}{M(B_m)}\right) + \left(\frac{M(N_m)}{M(X)} \cdot \frac{M_{m-1}}{M_n}\right)^2 \cdot \left(k_{N_m}^2 + \frac{K_{m-1}}{M(N_m)}\right) \right] \qquad Eq.A9$$

**Proof.** Because our random values are uncorrelated,

$$M(Y) = M(C_n(X)) + \sum_{m=1}^{n} [M(C_m(B_m)) + M(C_{m-1}(N_{m-1}))] \qquad Eq.\,A9.1$$

$$Var(Y) = Var(C_n(X)) + \sum_{m=1}^{n} [Var(C_m(B_m)) + Var(C_{m-1}(N_{m-1}))] \qquad Eq.\,A9.2$$

The equality for mean now follows from (Eq. A9.1) and (Eq. A4.1). To derive the equality for variance, we re-write (A4.2) as

$$Var(C_n(X)) = M(C_n(X))^2 \cdot \left( k_X^2 + \frac{K_n}{M(X)} \right)$$

and apply it and (Eq. A4.1) to all summands in the right side of (Eq. A9.2):

$$\frac{Var(Y)}{M(C_n(X))^2} = k_X^2 + \frac{K_n}{M(X)} + \Delta$$

where

$$\Delta = \frac{\sum_{m=1}^{n} \left[ (M(B_m) \cdot M_m)^2 \cdot \left( k_{B_m}^2 + \frac{K_m}{M(B_m)} \right) + (M(N_m) \cdot M_{m-1})^2 \cdot \left( k_{N_m}^2 + \frac{K_{m-1}}{M(N_m)} \right) \right]}{M_n^2 \cdot M(X)^2}$$

The last expression is clearly identical to eq. A9.